\documentclass{article}

\PassOptionsToPackage{numbers, compress}{natbib}
\usepackage[final]{neurips_2022}
\usepackage[symbol]{footmisc}

\usepackage[utf8]{inputenc} 
\usepackage[T1]{fontenc}    
\usepackage{hyperref}       
\usepackage{url}            
\usepackage{booktabs}       
\usepackage{amsfonts}       
\usepackage{nicefrac}       
\usepackage{microtype}      
\usepackage{xcolor}         
\usepackage{pdfcolfoot}
\usepackage{graphicx}

\usepackage{caption}
\usepackage{subcaption}
\begin{document}
\title{A Comparison of Speech Data Augmentation Methods Using S3PRL Toolkit}

%

\author{
  Mina Huh\thanks{Equal contribution.}\\
  Department of Computer Science\\
  University of Texas at Austin\\
  \texttt{minahuh@cs.utexas.edu} \\
   \And
   Ruchira Ray\footnotemark[1]\\
   Department of Computer Science\\
   University of Texas at Austin \\
   \texttt{ruchiraray@utexas.edu} \\
   \And
   Corey Karnei\footnotemark[1]\\
   Department of Electrical and Computer Engineering\\
   University of Texas at Austin \\
   \texttt{coreykarnei@utexas.edu}\\
}

\maketitle

\begin{abstract} 
Data augmentations are known to improve robustness in speech-processing tasks. In this study, we summarize and compare different data augmentation strategies using S3PRL toolkit. We explore how HuBERT and wav2vec perform using different augmentation techniques (SpecAugment, Gaussian Noise, Speed Perturbation) for Phoneme Recognition (PR) and Automatic Speech Recognition (ASR) tasks. We evaluate model performance in terms of phoneme error rate (PER) and word error rate (WER). From the experiments, we observed that SpecAugment slightly improves the performance of HuBERT and wav2vec on the original dataset. Also, we show that models trained using the Gaussian Noise and Speed Perturbation dataset are more robust when tested with augmented test sets. 
\end{abstract}

\section{Introduction} 
It is widely known that data augmentations can tackle the challenges of overfitting and improve generalizability for deep learning models ~\cite{park2019specaugment, ko2015audio, 2022improving}. In speech processing tasks, augmentations involve deforming the waveform by changing its speed, adding background noise, or applying augmentation policy to the audio spectrogram which is the image representation of the audio~\cite{park2019specaugment}. With these augmentations, the dataset is effectively enlarged as the multiple version of augmented data is fed instead of a single input. 

Phoneme Recognition (PR) and Automatic Speech Recognition (ASR) are two of the most common speech-processing tasks that can greatly benefit from applying these augmentation techniques. We used S3PRL, an open-source toolkit that targets self-supervised learning for speech processing ~\cite{yang2021superb}. It supports easy benchmarking of different speech representation models. In this paper, we chose to experiment with two pre-trained models, wav2vec~\cite{schneider2019wav2vec} and HuBERT~\cite{hsu2021hubert} to see how different augmentation techniques impact the performance of both PR and ASR tasks.

Our results show that applying SpecAugment in the data slightly improves the performance (PER and WER) on the raw test dataset and the augmented test set. We also demonstrate that experiments with augmented test datasets have the best results when the model was trained on that augmented training dataset. For the PR task, ~\textit{HuBERT-Gaussian-Noise} (13.10\%) and ~\textit{wav2vec-Gaussian-Noise} (70.67\%) showed the lowest PER on test sets with Gaussian Noise. For the ASR task, ~\textit{HuBERT-Speed-Perturbation} (21.63\%) and ~\textit{wav2vec-Speed-Perturbation} (34.22\%) showed the lowest WER on test sets with speed perturbation.  

\section{Background} 
\subsection{S3PRL} 
S3PRL is an open-source toolkit with powerful functionality at all levels of model training, development, and deployment. The name is an acronym for ~\textit{Self-Supervised Speech Pre-training and Representation Learning}. In this project, we utilize this toolkit for all of our model training and evaluation. 

There are three main components of the S3PRL toolkit. The first is ~\textit{Pre-Training}, which allows the user to train the models from scratch (\textit{e.g., } Mockingjay~\cite{liu2020mockingjay}, Audio ALBERT~\cite{chi2021audio}, TERA~\cite{liu2021tera}, etc). The second main component is ~\textit{Upstream models}, where the user can easily load more than 15 upstream models with pre-trained weights in a unified I/O interface. The final component is ~\textit{Downstream Evaluation}, where upstream models can be applied to a large number of downstream tasks and evaluated using the official implementation of the SUPERB Benchmark \cite{yang2021superb}.

\subsection{Augmentation Methods} 
End-to-end and hybrid (DNN-HMM) models are the two most common approaches for speech-based tasks like ASR~\cite{perero2022comparison, wang2019overview, li2022recent}. Their performance improves when there is a large enough amount of training data. However, these models excel at resource-intensive tasks while struggling in environments with limited resources.
There is a severe scarcity of reliable, diverse data for many tasks and low/resource languages~\cite{likhomanenko2020rethinking}. Overfitting models and inadequate generalization are the results of this. Data augmentation strategies have been shown to effectively reduce model overfitting.

Most acoustic models are built and adjusted on a single dataset and generalize poorly. In addition, the vast majority of big standard benchmarks share common characteristics in terms of domains and recording settings, such as low levels of background noise and reverberation. This causes studies to be conducted in isolated settings. Although there are benchmarks for noisy environments, their training sets are small \cite{likhomanenko2020rethinking, barker2018fifth, maciejewski2020whamr}.

Because of the temporal character of the spoken signal and the unique physical meaning of the spectrogram, the audio data has specific augmentation methods. These approaches, for example, are adding Gaussian Noise, temporal stretch, and pitch shift are three common augmentation methods for the raw audio stream. SpecAugment \cite{park2019specaugment} is a popular feature augmentation approach for spectrograms. The two-dimensional spectrogram diagram is interpreted as an image in this manner, with time on the horizontal axis and frequency on the vertical. This feature is augmented by time warping, frequency masking, and time masking. Some of the first studies of noise-tolerant ASR using deep networks were RNNs on Aurora-2 and DNNs on Aurora-4, respectively \cite{vinyals2012revisiting, seltzer2013investigation}. The first investigates the transfer performance of an RNN-based acoustic model trained purely on clean speech, while the second investigates alternative noise-aware training regimes for DNNs and which are most advantageous. The use of data augmentation on low-resource voice recognition tasks, in which models were assisted by generated data, is demonstrated in \cite{kanda2013elastic,xia2019generalized,ragni2014data}. In \cite{hannun2014deep}, in order to strengthen the model and make it more robust, a noisy audio source was layered on top of clear audio. Raw audio was manipulated in terms of speed in order to perform LVSCR tasks in \cite{ko2015audio}. However, cutting-edge models such as HuBERT \cite{hsu2021hubert} and wav2vec \cite{schneider2019wav2vec} (only trimming which was not used to expand the dataset) did not examine their models' performance with data augmentation. The available pre-trained models are trained with LibriSpeech, with no augmentation. In this work, we want to examine how finetuning with augmented data affects model robustness and generalization to real-world data (data with noise or change in speed).

In this research, we show that utilizing data augmentation not only increases the quantity of training data to avoid overfitting but also improves model robustness to real-world circumstances \cite{tokozume2017learning, park2019specaugment}. There are two types of audio augmentation: data augmentation and feature augmentation. Feature augmentation is concerned with augmenting data by altering the extracted features. SpecAugment \cite{park2019specaugment} is a popular technique for augmenting audio features. SpecAugment is concerned with Time Warping, Time Masking, and Frequency Masking. One of the shortcomings of the SpecAugment technique is that it does not introduce auditory diversity. Data augmentation is the process of enhancing a wave file without extracting any features. Typically, this includes changing the speed \cite{ko2015audio}, adding white noise \cite{eklund2019data}, adding background noise, changing the pitch, adding reverb, adding room impulse response, and so on.  


\section{Data}
\subsection{Dataset} 
In the experiment, we use LibriSpeech~\cite{7178964}, a publicly available corpus of read English speech that is suitable for training and evaluating multiple speech recognition systems. The dataset is originally derived from audiobooks from LibriVox\footnote{https://librivox.org/} project and contains 1,000 hours of speech samples (16 kHz).

\subsection{Data Preparation} 
Prior to applying any data augmentation method, we converted the audio file type from FLAC to WAV using FFmpeg~\cite{tomar2006converting}. Then, to enhance our training dataset, we used three different augmentation techniques (two for ASR task and two for PR task). We prepared models by training on different augmented training datasets and measured the performances using different augmented test datasets (refer to \autoref{tab:table_pr} and \autoref{tab:table_asr}).

\subsection{Data Augmentation} 

Here, we employ three different types of task-specific augmentation. We employ specaugment and gaussian noise for phoneme recognition \cite{moell2022speech}. For ASR, we employ SpecAugment and Speed Perturbation \cite{ko2015audio}. We adopt these enhancements since prior research has demonstrated their effectiveness for the aforementioned tasks. To augment the LibriSpeech data used in training, we used the open-source torchaudio library \cite{yang2021torchaudio}. Our goal in this data augmentation was to enhance the total number of samples while maintaining the dataset's balance and consistency.

We use Specaugment\cite{park2019specaugment} as our audio feature augmentation for both tasks. SpecAugment is a log mel spectrogram-based augmentation approach. SpecAugment fully exploits the features of the spectrogram to augment it. This study examines all three types of SpecAugment strategies: time warping, frequency masking, and time masking. Frequency masking uses a uniform distribution to apply a mask to number of consecutive mel frequency channels. Temporal masking, on the other hand, applies a mask to each subsequent time step and follows a uniform distribution. Masking enables the network to focus on the entire spectrogram as opposed to simply the time-frequency features of a certain frequency or period. Time warping involves warping a random point along a horizontal line passing through the center of the image during the time steps by a certain distance determined from a uniform distribution from 0 to the time warp parameter along that line.

With regards to Phoneme Recognization, we take into account additive White Gaussian Noise \cite{eklund2019data}. The signal can have this form of noise added to it using element-wise arithmetic addition. Additionally, it is drawn at random from a Gaussian distribution where the mean is zero. As a result, the standard deviation is not always the same. It has equal amounts of all frequency components. The RMS of the noise we need to produce can be calculated from the Signal Noise Ratio (SNR). In our project, we set SNR to 10. Then, we calculate the standard deviation from this. Random-number sampling from a Gaussian distribution with $mean=0$  and standard $deviation = {root\textrm{-}mean\textrm{-}squared}\:(RMS)$ is used to generate noise.

With regards to Phoneme Recognization, we take into account speed perturbation \cite{ko2015audio}. The process of speed perturbation includes speeding up or slowing down the audio without changing the pitch. We utilized torchaudio SoX \cite{yang2021torchaudio} to alter the training data's speed by factors of 0.9, 1.0, 1.1, 0.5, and 1.5, and then re-sampled the audio at the original rate. When the speed is altered, the resulting time signal is distorted. To illustrate, let's say we have an audio signal x(t), and we time warp it by a factor, and we get the signal x($\alpha$t). The frequency components are shifted by an amount proportional to the frequency due to the warping factor. This translates approximately to a shift of the spectrogram in the mel spectrogram. Speed Perturbation alters the signal's length and, consequently, the number of frames in the utterance. The SoX module of the torchaudio library was used for the speed perturbation technique. The SoX command applies specified effects on a tensor, such as speed and pitch change. We applied the speed effect while maintaining the same sampling rate. In order to add variety to the files, we built a random function that picked a factor to change the speed by at random.

\begin{figure}
\begin{subfigure}{.5\textwidth}
  \centering
  \includegraphics[width=.8\linewidth]{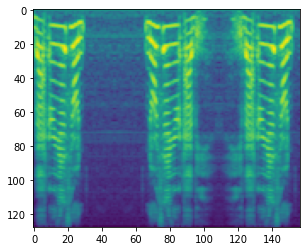}  
  \caption{Original}
  \label{fig:sub-first}
\end{subfigure}
\begin{subfigure}{.5\textwidth}
  \centering
  \includegraphics[width=.8\linewidth]{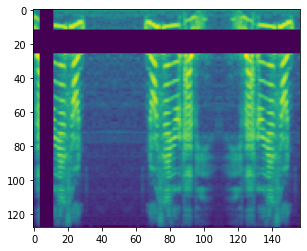}  
  \caption{SpecAugment}
  \label{fig:sub-second}
\end{subfigure}

\begin{subfigure}{.5\textwidth}
  \centering
  \includegraphics[width=.8\linewidth]{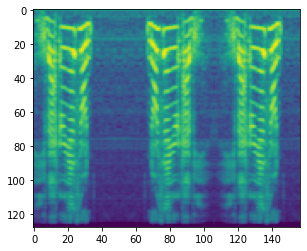}  
  \caption{Speed Perturbation}
  \label{fig:sub-third}
\end{subfigure}
\begin{subfigure}{.5\textwidth}
  \centering
  \includegraphics[width=.8\linewidth]{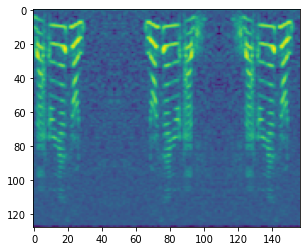}  
  \caption{Gaussian Noise}
  \label{fig:sub-fourth}
\end{subfigure}
\caption{\textit{Examples of spectrograms after data augmentation (Speed Perturbation and Gaussian Noise)((Fig \ref{fig:sub-third}, \ref{fig:sub-fourth}) is applied and feature augmentation (Fig \ref{fig:sub-second}) are applied separately. The original audio spectrogram is shown in the first spectrogram (Fig \ref{fig:sub-first})}}
\label{fig:fig}
\end{figure}

\section{Experiments} 

\begin{figure}[t]
\includegraphics[width=1\textwidth]{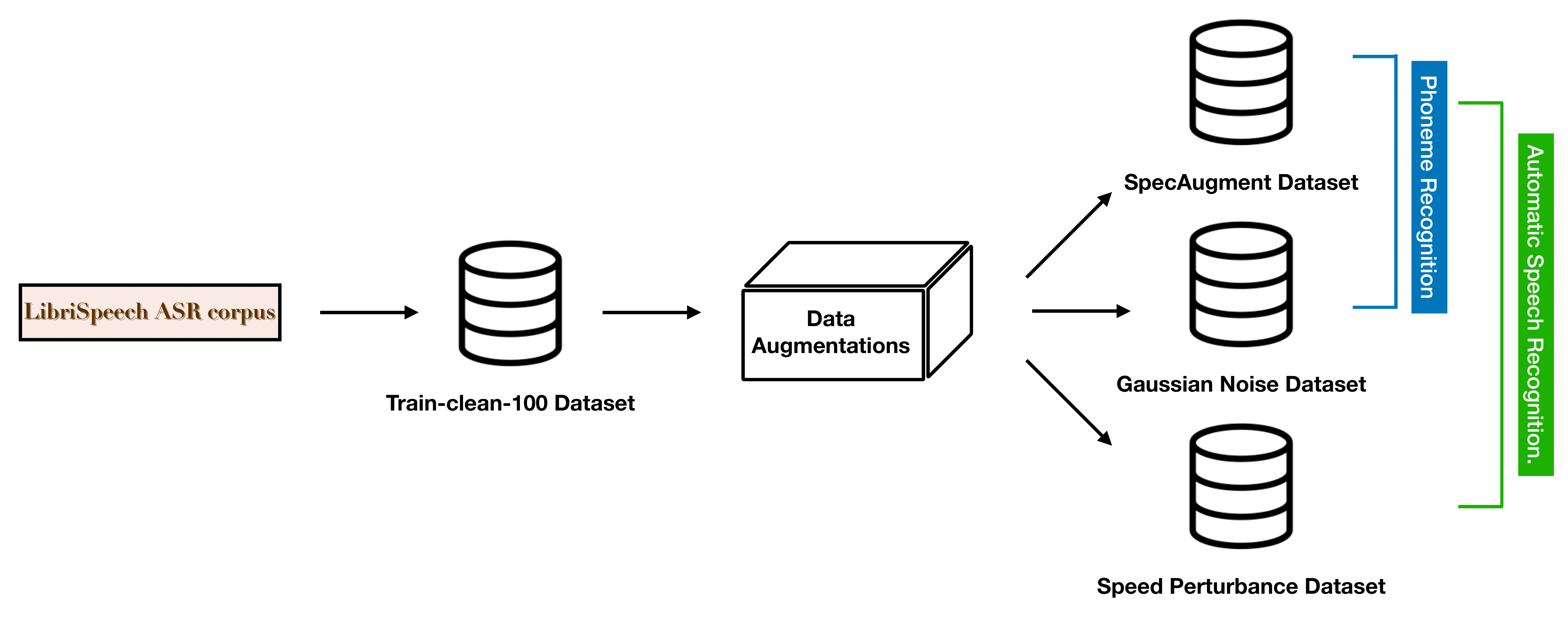}
\caption{\textit{We selected the 100-hour dataset from the LibriSpeech corpus and applied our chosen augmentations to it to generate an augmented dataset. For the Phoneme recognition task, the SpecAugment dataset and Gaussian Noise dataset were used. For the Automatic speech recognition task, the SpecAugment dataset and Speed Perturbation dataset were used.}}\label{fig:augmentation1}
\end{figure}

\subsection{Speech-Processing Tasks} 
General speech processing can be categorized into two tasks: discriminative and generative ~\cite{yang21c_interspeech}. The former discriminates continuous speech into discrete decisions (\textit{e.g.,} words in ASR, classes in speaker identification). The latter generates continuous speech
from inputs (\textit{e.g.,} TextTS, voice conversion). For this project, we chose two of the discriminative tasks: Phoneme Recognition (PR) and Automatic Speech Recognition (ASR).

PR is a speech recognition task that aims to transcribe an utterance into \textit{phonemes}. A phoneme is the smallest speech unit that can be perceptually distinguished among words in a language. Every word in a language can be considered as an ordered sequence of phonemes. SUPERB~\cite{yang2021superb} obtains the transcriptions of the phonemes from the LibriSpeech official g2p-model-5 and using the conversion script in Kaldi LibriSpeech
s5 recipe~\cite{Povey_ASRU2011}. For PR task, the evaluation metric used is the phone error rate (PER).

ASR is a task that transcribes utterances into \textit{words}. Also, known as Speech-to-text (STT), it transcribes the given audio to text having many real-world scenarios such as voice-user-interface. There are two main approaches for ASR: a traditional hybrid approach and a deep-learning-based approach. The components of the former approach include the lexicon model, acoustic model, and language model, where the acoustic model is usually of a Hidden-Markov Model (HMM) or a Gaussian Mixture Model (GMM) variant ~\cite{gales2008application}. The latter approach, end-to-end deep-learning models does not need to be force-aligned. The two main structures for end-to-end ASR are CTC~\cite{hannun2017sequence} and attention model~\cite{chorowski2015attention}. For the ASR task, the evaluation metric used is the word error rate (WER).

\subsection{Self-supervised Pre-trained Models} 
Among multiple self-supervised pre-trained models available by the S3PRL toolkit, we selected two of the discriminative modeling - wav2vec~\cite{schneider2019wav2vec} and HuBERT~\cite{hsu2021hubert}. 
\subsubsection{wav2vec}
Wav2Vec is a convolutional neural network (CNN) that takes raw audio as input and then computes a general representation. It proposed several architecture changes that have been shown to improve CPC ~\cite{riviere2020unsupervised}. vq-wav2vec~\cite{baevski2019vq} introduced the VQ model to wav2vec. It uses discrete speech representations, learning on a set of discrete values instead of continuous vectors. Wav2vec 2.0~\cite{baevski2020wav2vec} applied vq-wav2vec into but builds context representations over continuous speech representations. Its self-attention is capable of capturing dependencies over the end-to-end sequence of latent representations. Although later models such as wav2vec 2.0~\cite{baevski2020wav2vec} have shown better results on most of the speech processing tasks including ASR and PR, we wanted to explore our experiments on data augmentations and how they affect the model performance of wav2vec.

\subsubsection{HuBERT}
HuBERT is a self-supervised speech representation learning approach that utilizes offline clustering to enable a BERT's token prediction ~\cite{hsu2021hubert}. HuBERT overcomes the three common challenges in conventional self-supervised approaches: 1) having multiple sound units for each input utterances, 2) lacking input sound lexicon during the pre-training, and 3) having variable lengths of sound units without explicit segmentation. HuBERT Base and Large respectively outperform the wav2vec 2.0 Base and Large performances on both PR and ASR tasks.

\subsection{Experimental Setup} 
First, a task is selected; either phoneme recognition (PR) or automatic speech recognition (ASR). A HuBERT or Wav2Vec model is initialized with random weights and trained for 50,000 training steps on the original LibriSpeech dataset. HuBERT models are initialized with the Adam optimizer and a learning rate of 0.01 for PR and 0.0001 for ASR. Every 1000 steps during training a validation test is run. The current model is tested on a holdout validation set called "dev-clean" and if the error rate beats that of the previous best model, the current model weights are saved as the best model. 

Next, the best-saved model is fine-tuned three different times for another 50,000 training steps each on one of four datasets: Baseline (raw dataset), SpecAgument, Random Speed Perturbation, or Random Gaussian Noise. Since some of the augmentations are task-specific, not all datasets are used for each task. The ASR task is finetuned on the SpecAugment, and Speed Perturbation while the PR task is finetuned on SpecAgument, and Gaussian Noise. Conducting the experiment in this way ensures each model trains for the same amount of time while simulating the augmented models being trained on both the original and augmented dataset.

Finally, using the trained models, we test the performance of each task (PER for PR task, WER for ASR task) using test sets. Here, we test with both the original test set and augmented test set to explore the robustness of models under different scenarios. For PR task, each of the models is tested with both baseline and Gaussian Noise test set, and for ASR task, models are tested with both baseline and Speed Perturbation test set

\section{Results} 
%

\begin{table}[h]
    \centering
    \caption{\textit{Experimentation results for PR Task with different combinations of model and augmentation techniques. For the baseline dataset, we also report the PER presented in the original paper~\cite{yang2021superb}. The lowest PER value per each test dataset type is shown in bold.\\}}
\begin{tabular}{ |p{5cm}||p{3cm}|p{3cm}|  }
 \hline
 \multicolumn{3}{|c|}{Experimental Results on PR Task} \\
 \hline
 \multicolumn{1}{|c|}{{Model \& Augmentation Type}} & \multicolumn{2}{c|}{PER on the augmented test set (\%)}                           \\ 
 \cline{2-3} 
\multicolumn{1}{|l|}{}                      & \multicolumn{1}{c|}{Baseline (original)} & \multicolumn{1}{c|}{Gaussian Noise} \\ \hline
 \hline
 \textit{HuBERT-Baseline}   & 6.38 (5.41)    &  16.54\\
 \textit{HuBERT-SpecAugment} &   \textbf{6.11}  &  15.13  \\
 \textit{HuBERT-Gaussian-Noise} &13.17 & \textbf{13.10}\\
 \hline
 \textit{wav2vec-Baseline}     & \textbf{32.53} (31.58)   &  74.12\\
\textit{wav2vec-SpecAugment} &   32.60  &  72.15\\
\textit{wav2vec-Gaussian-Noise} &74.82 &  \textbf{70.67}\\
 \hline
\end{tabular}
    \label{tab:table_pr}
\end{table}

\begin{table}[h]
    \centering
    \caption{\textit{Experimentation results for ASR Task with different combinations of model and augmentation techniques. For the baseline dataset, we also report the WER presented in the original paper~\cite{yang2021superb}. The lowest WER value per each test dataset type is shown in bold.\\}}
\begin{tabular}{ |p{5cm}||p{3cm}|p{3cm}|  }
\hline
\multicolumn{3}{|c|}{Experimental Results on ASR Task} \\
\hline
 \multicolumn{1}{|c|}{{Model \& Augmentation Type}} & \multicolumn{2}{c|}{WER on the augmented test set (\%)}                           \\ \cline{2-3} 
\multicolumn{1}{|l|}{}                      & \multicolumn{1}{c|}{Baseline (original)} & \multicolumn{1}{c|}{Speed purturbation} \\ \hline
 \hline
 \textit{HuBERT-Baseline}    & 6.84 (6.42)& 30.36\\
 \textit{HuBERT-SpecAugment} &   \textbf{6.37}   & 29.13 \\
 \textit{HuBERT-Speed-Perturbation} &23.14  & \textbf{21.63} \\
 \hline 
 \textit{wav2vec-Baseline}    & 18.78 (15.86)  & 47.28\\
 \textit{wav2vec-SpecAugment} &   \textbf{17.31}   & 45.96\\
\textit{wav2vec-Speed-Perturbation}& 36.18 &  \textbf{34.22} \\
 \hline
\end{tabular}
    \label{tab:table_asr}
\end{table}

From the \autoref{tab:table_pr} and \autoref{tab:table_asr}, we can see that the reproduced PER and WER are not exactly the same as the ones presented in the original paper (values inside the parenthesis) ~\cite{yang2021superb}, which is potentially due to initial weight randomizations. 

We can see that SpecAugment is the only one of the augmentations that don't degrade model performance on the original dataset. Both the Gaussian Noise and Speed Perturbation augmentations decreased the performance of the HuBERT and wav2vec models when evaluated on the original dataset. When evaluated on their own respective datasets, however, their effects can be seen more clearly. Unsurprisingly, the models finetuned for the Gaussian Noise and Speed Perturbations perform better than the Baseline and SpecAugment models on their datasets respectively. These Gaussian Noise and Speed Perturbation models may not be more generalizable, but they are certainly more robust to the type of noise they're finetuned on compared to baseline and SpecAugment. SpecAugment's improvement was small, but the fact that it translates to performance improvement even on the Baseline dataset shows that it helps the model to learn a generalized distribution of the data.

\section{Discussion} 



In this research, we explored whether augmentation aids in making our models more robust to real-world data. \autoref{tab:table_pr} and \autoref{tab:table_asr} display the results of our research. We demonstrate that finetuning with noisy augmented data makes the model more resilient to such effects. \textit{HuBERT-Gaussian-Noise} reduces PER by more than 20\%, while \textit{HuBERT-Speed-Perturbation} reduces WER by more than 28\%. \textit{wav2vec-Gaussian-Noise} reduces PER by more than 4.6\%, whereas \textit{wav2vec-Speed-Perturbation} reduces WER by more than 27\%. We also notice \textit{HuBERT-SpecAugment} and \textit{wav2vec-SpecAugment} show lower PER and WER for both the original dataset and the augmented datasets. The performance of the models finetuned using data augmented by speed perturbation and Gaussian Noise was inferior on the original dataset. This is due to the fact that it shifts the learned distribution away from the original distribution. Specaugment, on the other hand, produces good results on the original data. This could be due to the fact that speculation does not produce a distribution change and does not introduce auditory diversity.

One limitation of our study is that we did not test our model on a variety of publically available datasets or on real-world data. We also did not test our models' outcomes on datasets augmented with SpecAugment. On audio augmented with real-world background noise, we expect our current finetuned model to outperform the original pre-trained model. 

For augmentation, there always remains the issue of generalization~\cite{xu2020wemix}. Previous research has shown that no single validation or test set produced from publicly available datasets is appropriate for quantifying transfer to other publicly available datasets or to real-world audio data \cite{likhomanenko2020rethinking}. From our experiments, we could see a similar issue occurs with augmented data, while the fine-tuned model performed better on noisy augmented data, it performed worse on the original data.

In the future, we plan to test our models on various publicly available datasets to see if they outperform the initial pre-trained models. We also intend to test our models in the wild by gathering a small custom dataset. We also want to explore how different models perform with such fine-tuning for different tasks. We also plan to finetune by a mixed augmentation \cite{wei2020comparison}, Room Impulse Response \cite{ko2017study}, and background noise \cite{hsu2019disentangling}.




\section{Conclusion} 
In this work, we have explored how different types of speech data augmentations can impact the robustness of pre-trained self-supervised models in discriminative tasks. Using the S3PRL toolkit, we explored how two upstream models (wav2vec and HuBERT) perform differently on two downstream tasks (PR and ASR) when trained using different types of augmented datasets (PR: raw dataset, SpecAugment dataset, and Gaussian Noise dataset, ASR: raw dataset, SpecAugment dataset, and Speed Perturbation dataset). Our results show that the SpecAugment technique has comparable results to the baseline model when experimented on the original test sets. We also observed that the models trained on noisy data (\textit{HuBERT-Gaussian-Noise} and \textit{wav2vec-Gaussian-Noise}) and models trained on speed-changed data (\textit{HuBERT-speed-perturbation} and \textit{wav2vec-speed-perturbation}) are more robust when tested with a more realistic data. 
As a future work, we plan to explore how sequentially augmented datasets (\textit{e.g., } SpecAugment + Gaussian Noise + Speed Perturbation) further increase the robustness. Also, we aim to evaluate the performance of these methods on different datasets than the LibriSpeech corpus, to claim the effectiveness with more in-the-wild data.

\small
\bibliographystyle{plainnat}
\bibliography{main}

\begin{thebibliography}{39}
\providecommand{\natexlab}[1]{#1}
\providecommand{\url}[1]{\texttt{#1}}
\expandafter\ifx\csname urlstyle\endcsname\relax
  \providecommand{\doi}[1]{doi: #1}\else
  \providecommand{\doi}{doi: \begingroup \urlstyle{rm}\Url}\fi

\bibitem[Baevski et~al.(2019)Baevski, Schneider, and Auli]{baevski2019vq}
Alexei Baevski, Steffen Schneider, and Michael Auli.
\newblock vq-wav2vec: Self-supervised learning of discrete speech
  representations.
\newblock \emph{arXiv preprint arXiv:1910.05453}, 2019.

\bibitem[Baevski et~al.(2020)Baevski, Zhou, Mohamed, and
  Auli]{baevski2020wav2vec}
Alexei Baevski, Yuhao Zhou, Abdelrahman Mohamed, and Michael Auli.
\newblock wav2vec 2.0: A framework for self-supervised learning of speech
  representations.
\newblock \emph{Advances in Neural Information Processing Systems},
  33:\penalty0 12449--12460, 2020.

\bibitem[Barker et~al.(2018)Barker, Watanabe, Vincent, and
  Trmal]{barker2018fifth}
Jon Barker, Shinji Watanabe, Emmanuel Vincent, and Jan Trmal.
\newblock The fifth'chime'speech separation and recognition challenge: dataset,
  task and baselines.
\newblock \emph{arXiv preprint arXiv:1803.10609}, 2018.

\bibitem[Chi et~al.(2021)Chi, Chung, Wu, Hsieh, Chen, Li, and
  Lee]{chi2021audio}
Po-Han Chi, Pei-Hung Chung, Tsung-Han Wu, Chun-Cheng Hsieh, Yen-Hao Chen,
  Shang-Wen Li, and Hung-yi Lee.
\newblock Audio albert: A lite bert for self-supervised learning of audio
  representation.
\newblock In \emph{2021 IEEE Spoken Language Technology Workshop (SLT)}, pages
  344--350. IEEE, 2021.

\bibitem[Chorowski et~al.(2015)Chorowski, Bahdanau, Serdyuk, Cho, and
  Bengio]{chorowski2015attention}
Jan~K Chorowski, Dzmitry Bahdanau, Dmitriy Serdyuk, Kyunghyun Cho, and Yoshua
  Bengio.
\newblock Attention-based models for speech recognition.
\newblock \emph{Advances in neural information processing systems}, 28, 2015.

\bibitem[Eklund(2019)]{eklund2019data}
Ville-Veikko Eklund.
\newblock Data augmentation techniques for robust audio analysis.
\newblock Master's thesis, Tempere University, 2019.

\bibitem[Gales et~al.(2008)Gales, Young, et~al.]{gales2008application}
Mark Gales, Steve Young, et~al.
\newblock The application of hidden markov models in speech recognition.
\newblock \emph{Foundations and Trends{\textregistered} in Signal Processing},
  1\penalty0 (3):\penalty0 195--304, 2008.

\bibitem[Hannun(2017)]{hannun2017sequence}
Awni Hannun.
\newblock Sequence modeling with ctc.
\newblock \emph{Distill}, 2\penalty0 (11):\penalty0 e8, 2017.

\bibitem[Hannun et~al.(2014)Hannun, Case, Casper, Catanzaro, Diamos, Elsen,
  Prenger, Satheesh, Sengupta, Coates, et~al.]{hannun2014deep}
Awni Hannun, Carl Case, Jared Casper, Bryan Catanzaro, Greg Diamos, Erich
  Elsen, Ryan Prenger, Sanjeev Satheesh, Shubho Sengupta, Adam Coates, et~al.
\newblock Deep speech: Scaling up end-to-end speech recognition.
\newblock \emph{arXiv preprint arXiv:1412.5567}, 2014.

\bibitem[Hsu et~al.(2019)Hsu, Zhang, Weiss, Chung, Wang, Wu, and
  Glass]{hsu2019disentangling}
Wei-Ning Hsu, Yu~Zhang, Ron~J Weiss, Yu-An Chung, Yuxuan Wang, Yonghui Wu, and
  James Glass.
\newblock Disentangling correlated speaker and noise for speech synthesis via
  data augmentation and adversarial factorization.
\newblock In \emph{ICASSP 2019-2019 IEEE International Conference on Acoustics,
  Speech and Signal Processing (ICASSP)}, pages 5901--5905. IEEE, 2019.

\bibitem[Hsu et~al.(2021)Hsu, Bolte, Tsai, Lakhotia, Salakhutdinov, and
  Mohamed]{hsu2021hubert}
Wei-Ning Hsu, Benjamin Bolte, Yao-Hung~Hubert Tsai, Kushal Lakhotia, Ruslan
  Salakhutdinov, and Abdelrahman Mohamed.
\newblock Hubert: Self-supervised speech representation learning by masked
  prediction of hidden units.
\newblock \emph{IEEE/ACM Transactions on Audio, Speech, and Language
  Processing}, 29:\penalty0 3451--3460, 2021.

\bibitem[Kanda et~al.(2013)Kanda, Takeda, and Obuchi]{kanda2013elastic}
Naoyuki Kanda, Ryu Takeda, and Yasunari Obuchi.
\newblock Elastic spectral distortion for low resource speech recognition with
  deep neural networks.
\newblock In \emph{2013 IEEE Workshop on Automatic Speech Recognition and
  Understanding}, pages 309--314. IEEE, 2013.

\bibitem[Ko et~al.(2015)Ko, Peddinti, Povey, and Khudanpur]{ko2015audio}
Tom Ko, Vijayaditya Peddinti, Daniel Povey, and Sanjeev Khudanpur.
\newblock Audio augmentation for speech recognition.
\newblock In \emph{Sixteenth annual conference of the international speech
  communication association}, 2015.

\bibitem[Ko et~al.(2017)Ko, Peddinti, Povey, Seltzer, and
  Khudanpur]{ko2017study}
Tom Ko, Vijayaditya Peddinti, Daniel Povey, Michael~L Seltzer, and Sanjeev
  Khudanpur.
\newblock A study on data augmentation of reverberant speech for robust speech
  recognition.
\newblock In \emph{2017 IEEE International Conference on Acoustics, Speech and
  Signal Processing (ICASSP)}, pages 5220--5224. IEEE, 2017.

\bibitem[Li et~al.(2022)]{li2022recent}
Jinyu Li et~al.
\newblock Recent advances in end-to-end automatic speech recognition.
\newblock \emph{APSIPA Transactions on Signal and Information Processing},
  11\penalty0 (1), 2022.

\bibitem[Likhomanenko et~al.(2020)Likhomanenko, Xu, Pratap, Tomasello, Kahn,
  Avidov, Collobert, and Synnaeve]{likhomanenko2020rethinking}
Tatiana Likhomanenko, Qiantong Xu, Vineel Pratap, Paden Tomasello, Jacob Kahn,
  Gilad Avidov, Ronan Collobert, and Gabriel Synnaeve.
\newblock Rethinking evaluation in asr: Are our models robust enough?
\newblock \emph{arXiv preprint arXiv:2010.11745}, 2020.

\bibitem[Liu et~al.(2020)Liu, Yang, Chi, Hsu, and Lee]{liu2020mockingjay}
Andy~T Liu, Shu-wen Yang, Po-Han Chi, Po-chun Hsu, and Hung-yi Lee.
\newblock Mockingjay: Unsupervised speech representation learning with deep
  bidirectional transformer encoders.
\newblock In \emph{ICASSP 2020-2020 IEEE International Conference on Acoustics,
  Speech and Signal Processing (ICASSP)}, pages 6419--6423. IEEE, 2020.

\bibitem[Liu et~al.(2021)Liu, Li, and Lee]{liu2021tera}
Andy~T Liu, Shang-Wen Li, and Hung-yi Lee.
\newblock Tera: Self-supervised learning of transformer encoder representation
  for speech.
\newblock \emph{IEEE/ACM Transactions on Audio, Speech, and Language
  Processing}, 29:\penalty0 2351--2366, 2021.

\bibitem[Maciejewski et~al.(2020)Maciejewski, Wichern, McQuinn, and
  Le~Roux]{maciejewski2020whamr}
Matthew Maciejewski, Gordon Wichern, Emmett McQuinn, and Jonathan Le~Roux.
\newblock Whamr!: Noisy and reverberant single-channel speech separation.
\newblock In \emph{ICASSP 2020-2020 IEEE International Conference on Acoustics,
  Speech and Signal Processing (ICASSP)}, pages 696--700. IEEE, 2020.

\bibitem[Mo{\"e}ll et~al.(2022)Mo{\"e}ll, O'Regan, Mehta, Kirkland, Lameris,
  Gustafsson, and Beskow]{moell2022speech}
Birger Mo{\"e}ll, Jim O'Regan, Shivam Mehta, Ambika Kirkland, Harm Lameris,
  Joakim Gustafsson, and Jonas Beskow.
\newblock Speech data augmentation for improving phoneme transcriptions of
  aphasic speech using wav2vec 2.0 for the psst challenge.
\newblock In \emph{13th Language Resources and Evaluation Conference (LREC)},
  pages 62--70, 2022.

\bibitem[Oneaț{\u{a}} and Cucu(2022)]{2022improving}
Dan Oneaț{\u{a}} and Horia Cucu.
\newblock Improving multimodal speech recognition by data augmentation and
  speech representations.
\newblock In \emph{Proceedings of the IEEE/CVF Conference on Computer Vision
  and Pattern Recognition}, pages 4579--4588, 2022.

\bibitem[Panayotov et~al.(2015)Panayotov, Chen, Povey, and Khudanpur]{7178964}
Vassil Panayotov, Guoguo Chen, Daniel Povey, and Sanjeev Khudanpur.
\newblock Librispeech: An asr corpus based on public domain audio books.
\newblock In \emph{2015 IEEE International Conference on Acoustics, Speech and
  Signal Processing (ICASSP)}, pages 5206--5210, 2015.
\newblock \doi{10.1109/ICASSP.2015.7178964}.

\bibitem[Park et~al.(2019)Park, Chan, Zhang, Chiu, Zoph, Cubuk, and
  Le]{park2019specaugment}
Daniel~S Park, William Chan, Yu~Zhang, Chung-Cheng Chiu, Barret Zoph, Ekin~D
  Cubuk, and Quoc~V Le.
\newblock Specaugment: A simple data augmentation method for automatic speech
  recognition.
\newblock \emph{arXiv preprint arXiv:1904.08779}, 2019.

\bibitem[Perero-Codosero et~al.(2022)Perero-Codosero, Espinoza-Cuadros, and
  Hern{\'a}ndez-G{\'o}mez]{perero2022comparison}
Juan~M Perero-Codosero, Fernando~M Espinoza-Cuadros, and Luis~A
  Hern{\'a}ndez-G{\'o}mez.
\newblock A comparison of hybrid and end-to-end asr systems for the
  iberspeech-rtve 2020 speech-to-text transcription challenge.
\newblock \emph{Applied Sciences}, 12\penalty0 (2):\penalty0 903, 2022.

\bibitem[Povey et~al.(2011)Povey, Ghoshal, Boulianne, Burget, Glembek, Goel,
  Hannemann, Motlicek, Qian, Schwarz, Silovsky, Stemmer, and
  Vesely]{Povey_ASRU2011}
Daniel Povey, Arnab Ghoshal, Gilles Boulianne, Lukas Burget, Ondrej Glembek,
  Nagendra Goel, Mirko Hannemann, Petr Motlicek, Yanmin Qian, Petr Schwarz, Jan
  Silovsky, Georg Stemmer, and Karel Vesely.
\newblock The kaldi speech recognition toolkit.
\newblock In \emph{IEEE 2011 Workshop on Automatic Speech Recognition and
  Understanding}. IEEE Signal Processing Society, December 2011.
\newblock IEEE Catalog No.: CFP11SRW-USB.

\bibitem[Ragni et~al.(2014)Ragni, Knill, Rath, and Gales]{ragni2014data}
Anton Ragni, Kate~M Knill, Shakti~P Rath, and Mark~JF Gales.
\newblock Data augmentation for low resource languages.
\newblock In \emph{INTERSPEECH 2014: 15th Annual Conference of the
  International Speech Communication Association}, pages 810--814.
  International Speech Communication Association (ISCA), 2014.

\bibitem[Riviere et~al.(2020)Riviere, Joulin, Mazar{\'e}, and
  Dupoux]{riviere2020unsupervised}
Morgane Riviere, Armand Joulin, Pierre-Emmanuel Mazar{\'e}, and Emmanuel
  Dupoux.
\newblock Unsupervised pretraining transfers well across languages.
\newblock In \emph{ICASSP 2020-2020 IEEE International Conference on Acoustics,
  Speech and Signal Processing (ICASSP)}, pages 7414--7418. IEEE, 2020.

\bibitem[Schneider et~al.(2019)Schneider, Baevski, Collobert, and
  Auli]{schneider2019wav2vec}
Steffen Schneider, Alexei Baevski, Ronan Collobert, and Michael Auli.
\newblock wav2vec: Unsupervised pre-training for speech recognition.
\newblock \emph{arXiv preprint arXiv:1904.05862}, 2019.

\bibitem[Seltzer et~al.(2013)Seltzer, Yu, and Wang]{seltzer2013investigation}
Michael~L Seltzer, Dong Yu, and Yongqiang Wang.
\newblock An investigation of deep neural networks for noise robust speech
  recognition.
\newblock In \emph{2013 IEEE international conference on acoustics, speech and
  signal processing}, pages 7398--7402. IEEE, 2013.

\bibitem[Tokozume et~al.(2017)Tokozume, Ushiku, and
  Harada]{tokozume2017learning}
Yuji Tokozume, Yoshitaka Ushiku, and Tatsuya Harada.
\newblock Learning from between-class examples for deep sound recognition.
\newblock \emph{arXiv preprint arXiv:1711.10282}, 2017.

\bibitem[Tomar(2006)]{tomar2006converting}
Suramya Tomar.
\newblock Converting video formats with ffmpeg.
\newblock \emph{Linux Journal}, 2006\penalty0 (146):\penalty0 10, 2006.

\bibitem[Vinyals et~al.(2012)Vinyals, Ravuri, and Povey]{vinyals2012revisiting}
Oriol Vinyals, Suman~V Ravuri, and Daniel Povey.
\newblock Revisiting recurrent neural networks for robust asr.
\newblock In \emph{2012 IEEE international conference on acoustics, speech and
  signal processing (ICASSP)}, pages 4085--4088. IEEE, 2012.

\bibitem[Wang et~al.(2019)Wang, Wang, and Lv]{wang2019overview}
Dong Wang, Xiaodong Wang, and Shaohe Lv.
\newblock An overview of end-to-end automatic speech recognition.
\newblock \emph{Symmetry}, 11\penalty0 (8):\penalty0 1018, 2019.

\bibitem[Wei et~al.(2020)Wei, Zou, Liao, et~al.]{wei2020comparison}
Shengyun Wei, Shun Zou, Feifan Liao, et~al.
\newblock A comparison on data augmentation methods based on deep learning for
  audio classification.
\newblock In \emph{Journal of Physics: Conference Series}, volume 1453, page
  012085. IOP Publishing, 2020.

\bibitem[wen Yang et~al.(2021)wen Yang, Chi, Chuang, Lai, Lakhotia, Lin, Liu,
  Shi, Chang, Lin, Huang, Tseng, tik Lee, Liu, Huang, Dong, Li, Watanabe,
  Mohamed, and yi~Lee]{yang21c_interspeech}
Shu wen Yang, Po-Han Chi, Yung-Sung Chuang, Cheng-I~Jeff Lai, Kushal Lakhotia,
  Yist~Y. Lin, Andy~T. Liu, Jiatong Shi, Xuankai Chang, Guan-Ting Lin,
  Tzu-Hsien Huang, Wei-Cheng Tseng, Ko~tik Lee, Da-Rong Liu, Zili Huang, Shuyan
  Dong, Shang-Wen Li, Shinji Watanabe, Abdelrahman Mohamed, and Hung yi~Lee.
\newblock {SUPERB: Speech Processing Universal PERformance Benchmark}.
\newblock In \emph{Proc. Interspeech 2021}, pages 1194--1198, 2021.
\newblock \doi{10.21437/Interspeech.2021-1775}.

\bibitem[Xia et~al.(2019)Xia, Kong, Anastasopoulos, and
  Neubig]{xia2019generalized}
Mengzhou Xia, Xiang Kong, Antonios Anastasopoulos, and Graham Neubig.
\newblock Generalized data augmentation for low-resource translation.
\newblock \emph{arXiv preprint arXiv:1906.03785}, 2019.

\bibitem[Xu et~al.(2020)Xu, Noy, Lin, Qian, Li, and Jin]{xu2020wemix}
Yi~Xu, Asaf Noy, Ming Lin, Qi~Qian, Hao Li, and Rong Jin.
\newblock Wemix: How to better utilize data augmentation.
\newblock \emph{arXiv preprint arXiv:2010.01267}, 2020.

\bibitem[Yang et~al.(2021{\natexlab{a}})Yang, Chi, Chuang, Lai, Lakhotia, Lin,
  Liu, Shi, Chang, Lin, et~al.]{yang2021superb}
Shu-wen Yang, Po-Han Chi, Yung-Sung Chuang, Cheng-I~Jeff Lai, Kushal Lakhotia,
  Yist~Y Lin, Andy~T Liu, Jiatong Shi, Xuankai Chang, Guan-Ting Lin, et~al.
\newblock Superb: Speech processing universal performance benchmark.
\newblock \emph{arXiv preprint arXiv:2105.01051}, 2021{\natexlab{a}}.

\bibitem[Yang et~al.(2021{\natexlab{b}})Yang, Hira, Ni, Chourdia, Astafurov,
  Chen, Yeh, Puhrsch, Pollack, Genzel, Greenberg, Yang, Lian, Mahadeokar,
  Hwang, Chen, Goldsborough, Roy, Narenthiran, Watanabe, Chintala,
  Quenneville-Bélair, and Shi]{yang2021torchaudio}
Yao-Yuan Yang, Moto Hira, Zhaoheng Ni, Anjali Chourdia, Artyom Astafurov,
  Caroline Chen, Ching-Feng Yeh, Christian Puhrsch, David Pollack, Dmitriy
  Genzel, Donny Greenberg, Edward~Z. Yang, Jason Lian, Jay Mahadeokar, Jeff
  Hwang, Ji~Chen, Peter Goldsborough, Prabhat Roy, Sean Narenthiran, Shinji
  Watanabe, Soumith Chintala, Vincent Quenneville-Bélair, and Yangyang Shi.
\newblock Torchaudio: Building blocks for audio and speech processing.
\newblock \emph{arXiv preprint arXiv:2110.15018}, 2021{\natexlab{b}}.

\end{thebibliography}

\end{document}